\documentclass[12pt]{article}

\usepackage{amsmath}
\usepackage{amssymb}
\usepackage{epsfig}
\usepackage{graphicx}
\usepackage{cite}
\usepackage{multirow}
\usepackage{longtable}
\usepackage{lscape}
\usepackage{bbm}
\usepackage{dcolumn}
\usepackage[dvips]{color}
\usepackage{rotating}

\usepackage{url}

\allowdisplaybreaks[4] 

\newcommand{\capdef}{}
\newcommand{\mycaption}[2][\capdef]{\renewcommand{\capdef}{#2}%
        \caption[#1]{{\footnotesize #2}}}
\makeatletter
\renewcommand{\fnum@table}{\textbf{\tablename~\thetable}}
\renewcommand{\fnum@figure}{\textbf{\figurename~\thefigure}}
\makeatother

\newcounter{myenumi}

\renewcommand{\themyenumi}{\roman{myenumi}}
{\end{list}}

\newlength{\myem}
\newcounter{mysubequation}[equation]

\makeatletter
\renewcommand{\section}{\@startsection{section}{1}{0em}{-\baselineskip}%
{\baselineskip}{\normalfont\large\bfseries}}
\renewcommand{\subsection}%
{\@startsection{subsection}{2}{0em}{-0.7\baselineskip}%
{0.7\baselineskip}{\normalfont\bfseries}}
\makeatother

\newcommand{\bi}{\begin{itemize}}
\newcommand{\ei}{\end{itemize}}

\newcommand{\be}{\begin{equation}}
\newcommand{\ee}{\end{equation}}
\newcommand{\bea}{\begin{eqnarray}}
\newcommand{\eea}{\end{eqnarray}}

\newcommand{\ie}{{\it i.e.}}

\newcommand{\eg}{{\it e.g.}}

\newcommand{\cf}{{\it cf.}}

\newcommand{\etc}{{\it etc.}}
\newcommand{\eq}{Eq.}

\newcommand{\fig}{Fig.}

\newcommand{\Ref}{Ref.}
\newcommand{\Refs}{Refs.}

\newcommand{\equ}[1]{\eq~(\ref{equ:#1})}
\newcommand{\figu}[1]{\fig~\ref{fig:#1}}

\begin{document}

\begin{titlepage}

\renewcommand{\thefootnote}{\alph{footnote}}

\vspace*{-3.cm}
\begin{flushright}
EURONU-WP6-09-07
\end{flushright}


\renewcommand{\thefootnote}{\fnsymbol{footnote}}
\setcounter{footnote}{-1}

{\begin{center}
{\large\bf
Short-Baseline Electron Neutrino Disappearance \\ at a Neutrino Factory
} 
\end{center}}

\renewcommand{\thefootnote}{\alph{footnote}}

\vspace*{.8cm}
\vspace*{.3cm}
{\begin{center} {\large{\sc
                Carlo~Giunti\footnote[1]{\makebox[1.cm]{Email:}
                giunti@to.infn.it},
                Marco~Laveder\footnote[2]{\makebox[1.cm]{Email:}
                laveder@pd.infn.it}, and
                Walter~Winter\footnote[3]{\makebox[1.cm]{Email:}
                winter@physik.uni-wuerzburg.de}
                }}
\end{center}}
\vspace*{0cm}
{\it
\begin{center}

\footnotemark[1]
INFN, Sezione di Torino,
Via P. Giuria 1, I--10125 Torino, Italy

\footnotemark[2]
Dipartimento di Fisica ``G. Galilei'', Universit\`a di Padova
and
\\
INFN, Sezione di Padova,
Via F. Marzolo 8, I--35131 Padova, Italy

\footnotemark[3]
       Institut f{\"u}r Theoretische Physik und Astrophysik, \\ Universit{\"a}t W{\"u}rzburg, 
       D--97074 W{\"u}rzburg, Germany

\end{center}}

\vspace*{1.5cm}

\begin{center}
{\Large 18 September 2009}
\end{center}

{\Large \bf
\begin{center} Abstract \end{center}  }

We discuss short-baseline and very-short-baseline $\nu_e$ disappearance at a neutrino factory.
We take into account geometric effects, such as from averaging over the decay straights, and the uncertainties of the cross sections.
We follow an approach similar to reactor experiments with two detectors: we use two sets of near detectors at different distances to cancel systematics.
We demonstrate that such a setup is very robust with respect to systematics, and can have excellent sensitivities to the effective mixing angle and squared-mass splitting.
In addition, we allow for CPT invariance violation, which can be tested (depending on the parameters) up to a 0.1\% level.

\vspace*{.5cm}

\end{titlepage}

\newpage

\renewcommand{\thefootnote}{\arabic{footnote}}
\setcounter{footnote}{0}

\section{Introduction}
\label{Introduction}

Neutrino oscillation experiments have shown that neutrinos are massive particles with at least two
squared-mass differences:
$
\Delta{m}^{2}_{\text{SOL}}
\simeq 8 \times 10^{-5} \, \text{eV}^{2}
$,
measured in solar and very-long-baseline reactor neutrino experiments,
and
$
\Delta{m}^{2}_{\text{ATM}}
\simeq 2 \times 10^{-3} \, \text{eV}^{2}
$,
measured in atmospheric and long-baseline neutrino experiments
(see Refs.~\cite{hep-ph/9812360,hep-ph/0202058,hep-ph/0310238,hep-ph/0405172,hep-ph/0506083,hep-ph/0606054,Giunti-Kim-2007,GonzalezGarcia:2007ib,0805.2517,0808.2016}).
These two $\Delta{m}^{2}$'s are perfectly accommodated in the framework of
three-neutrino mixing, where there are two independent squared-mass differences.
However, there are experimental anomalies which may indicate the existence of
Short-BaseLine (SBL) or
Very-Short-BaseLine (VSBL) oscillations generated by a third
$\Delta{m}^{2}$ which is much larger than the other two:
$ \Delta{m}^{2}_{\text{SBL}} \gtrsim 10^{-1} \, \text{eV}^2 $
or
$ \Delta{m}^{2}_{\text{VSBL}} \gtrsim 10 \, \text{eV}^2 $.
Among these anomalies,
the most well-known is the LSND signal in favor of SBL $\bar\nu_{\mu}\to\bar\nu_{e}$ oscillations \cite{hep-ex/0104049},
which has not been confirmed by other experiments
and is currently disfavored by the negative results of
KARMEN \cite{hep-ex/0203021}
and
MiniBooNE \cite{AguilarArevalo:2008rc}.
Less well-known are the Gallium radioactive source experiments anomaly \cite{Abdurashitov:2009tn}
and
the MiniBooNE low-energy anomaly \cite{AguilarArevalo:2008rc},
which could be explained by
SBL \cite{Giunti:2006bj,Acero:2007su}
or
VSBL \cite{Giunti:2007xv,Giunti:2009zz} $\nu_e$ disappearance.

The existence of a third $\Delta{m}^{2}$ requires the existence of at least a fourth massive neutrino
which corresponds,
in the flavor basis,
to the existence of a sterile neutrino $\nu_{s}$,
\ie, a fermion which is a singlet under the Standard Model symmetries.
Hence it is electrically neutral and does not take part in weak interactions.
If the three active neutrinos
$\nu_{e}$,
$\nu_{\mu}$, and
$\nu_{\tau}$
are mixed with the sterile neutrino,
neutrino oscillation experiments can observe the disappearance of active neutrinos
into $\nu_{s}$.

In light of the above-mentioned anomalies,
it is interesting to investigate the possibility of (V)SBL $\nu_{e}$ disappearance
with future high-precision experiments.
In general,
it is important to investigate the possibility of $\nu_{e}$ disappearance
generated by a $\Delta{m}^{2}$ different from
$\Delta{m}^{2}_{\text{SOL}}$
and
$\Delta{m}^{2}_{\text{ATM}}$
in order to constrain schemes with mixing of four
(see Refs.~\cite{hep-ph/9812360,hep-ph/0405172,hep-ph/0606054,GonzalezGarcia:2007ib})
or more
\cite{hep-ph/0305255,0906.1997}
massive neutrinos.
These schemes have been studied mostly in connection with the LSND anomaly,
but the latest global fits of the experimental data, including the LSND signal,
are not good
\cite{hep-ph/0405172,GonzalezGarcia:2007ib}.
However, the schemes with mixing of more than three neutrinos
may be realized in nature independently of the LSND signal.
Hence, it is important to investigate the phenomenology of
sterile neutrinos with an open mind,
not only through neutrino oscillations
\cite{hep-ph/0609177,hep-ph/0611178,hep-ex/0701004,0704.0388,0705.0107,0706.1462,0707.2481,0710.2985,0907.3145},
but also by studying their effects in
astrophysics
\cite{0706.0399,0709.1937,0710.5180,0712.1816,0805.4014,0806.3029}
and cosmology
\cite{0711.2450,0810.5133,0812.2249}.

If there is (V)SBL electron neutrino disappearance,
it must be mainly into sterile neutrinos,
because the mixing of the three active neutrinos with the fourth massive neutrino
must be small in order to fit the data on $\nu_{e}\to\nu_{\mu,\tau}$ oscillations
generated by
$\Delta{m}^{2}_{\text{SOL}}$
and the data on $\nu_{\mu}\to\nu_{\tau}$ oscillations
generated by
$\Delta{m}^{2}_{\text{ATM}}$.
In the 3+1 four-neutrino schemes
(see Refs.~\cite{hep-ph/9812360,hep-ph/0405172,hep-ph/0606054,GonzalezGarcia:2007ib})
with
$ \Delta{m}^{2}_{\text{(V)SBL}} = |\Delta{m}^2_{41}| \gg \Delta{m}^{2}_{\text{ATM}} = |\Delta{m}^2_{31}| \gg \Delta{m}^{2}_{\text{SOL}} = |\Delta{m}^2_{21}| $,
where
$ \Delta{m}^2_{kj} \equiv m_{k}^2 - m_{j}^2 $,
the mixing matrix $U$
must be such that
$ |U_{e4}|, |U_{\mu4}|, |U_{\tau4}| \ll 1 $
and
$ |U_{s4}| \simeq 1 $.
Therefore,
the amplitudes of the (V)SBL oscillation channels,
$ A_{\alpha\beta} = 4 | U_{\alpha4} |^2 | U_{\beta4} |^2 $
for $\alpha\neq\beta$,
are such that
$ A_{ab} \ll A_{as} $
for
$a,b=e,\mu,\tau$.

In this paper we study the sensitivity of neutrino factory experiments
to (V)SBL $\nu_{e}$ and $\bar\nu_{e}$ disappearance,
which in practice has been investigated so far mainly through
SBL reactor neutrino experiments ($\bar\nu_{e}$ disappearance).

We will first study,
in Section~\ref{disappearance},
(V)SBL $\nu_{e}$ and $\bar\nu_{e}$ disappearance
at a neutrino factory assuming exact CPT symmetry,
which implies
$ P_{ee} = P_{\bar e \bar e} $
(see Ref.~\cite{Giunti-Kim-2007}),
considering the simplest case of effective two-neutrino mixing
with
\begin{equation}
P_{ee} = P_{\bar e \bar e}
=
1 - \sin^2 (2 \theta) \, \sin^2 \left( \frac{\Delta m^2 L}{4 E} \right)
\,,
\label{pee-cpt}
\end{equation}
where,
from now on,
$ \Delta{m}^{2} = \Delta{m}^{2}_{\text{(V)SBL}} $.
This is the case of four-neutrino mixing schemes with
$ \Delta{m}^{2} = |\Delta{m}^2_{41}| \gg \Delta{m}^{2}_{\text{ATM}} = |\Delta{m}^2_{31}| \gg \Delta{m}^{2}_{\text{SOL}} = |\Delta{m}^2_{21}| $.
In the 3+1 schemes,
the amplitude of the oscillations is related to the $U_{e4}$ element of the mixing matrix by
$ \sin^2 (2 \theta) = 4 |U_{e4}|^2 \left( 1 - |U_{e4}|^2  \right) $
(see Refs.~\cite{hep-ph/9812360,hep-ph/0405172,hep-ph/0606054,GonzalezGarcia:2007ib}).

The CPT symmetry is widely believed to be exact,
because it is a fundamental symmetry of local relativistic Quantum Field Theory
(see Ref.~\cite{hep-ph/0309309}).
However,
in recent years studies of extensions of the Standard Model
have shown that it is possible to have violations of the Lorentz
and CPT symmetries (see Refs.~\cite{hep-ph/0201258,hep-ph/0203261,0801.0287})
and
several phenomenological studies of neutrino oscillations with different
masses and mixing for neutrinos and antineutrinos
appeared in the literature
\cite{hep-ph/0010178,hep-ph/0108199,hep-ph/0112226,hep-ph/0201080,hep-ph/0201134,hep-ph/0201211,hep-ph/0307127,hep-ph/0308299,hep-ph/0505133,hep-ph/0306226,0804.2820,0903.4318}.
We will consider this scenario in the simplest case of effective two-neutrino mixing
with
\begin{eqnarray}
P_{ee} & = & 1 - \sin^2 (2 \theta_\nu) \, \sin^2 \left( \frac{\Delta m_\nu^2 L}{4 E} \right)
\,,
\label{equ:pelel}
\\
P_{\bar e \bar e} & = & 1 - \sin^2 (2 \theta_{\bar \nu}) \, \sin^2 \left( \frac{\Delta m_{\bar \nu}^2 L}{4 E} \right)
\,.
\label{equ:paeae}
\end{eqnarray}
This kind of CPT violation in a four-neutrino mixing scheme could reconcile the LSND signal with the other neutrino oscillation data
\cite{hep-ph/0308299}
and/or
could explain
the Gallium radioactive source experiments anomaly
and
the MiniBooNE low-energy anomaly
together with the absence of $\bar\nu_{e}$ disappearance in
reactor neutrino experiments \cite{Giunti:2009zz}.
Let us emphasize that the reconciliation of the LSND anomaly with the results of other neutrino oscillation experiments
is not possible in three-neutrino mixing schemes even if CPT violation is allowed
\cite{hep-ph/0306226,GonzalezGarcia:2007ib}.

Another hint in favor of a possible CPT violation comes from the
recent measurement of $\nu_{\mu}$ and $\bar\nu_{\mu}$ disappearance in the MINOS experiment
\cite{Evans-HEP2009},
which indicate different best-fit values of the oscillation parameters of
$\nu_{\mu}$ and $\bar\nu_{\mu}$:
$ \Delta\overline{m}^2_{\text{MINOS}} \simeq 2 \times 10^{-2} \, \text{eV}^2 $
and
$ \sin^2\overline{\theta}_{\text{MINOS}} \simeq 0.6 $
for $\bar\nu_{\mu}$'s,
whereas
$ \Delta{m}^2_{\text{MINOS}} \simeq 2.4 \times 10^{-3} \, \text{eV}^2 $
and
$ \sin^2\theta_{\text{MINOS}} \simeq 1 $
for $\nu_{\mu}$'s.
The best-fit values and allowed region of the $\nu_{\mu}$ oscillation parameters
are in agreement with atmospheric $\nu_{\mu}\to\nu_{\tau}$ oscillations.
Since the 90\% C.L. allowed region of the $\bar\nu_{\mu}$ oscillation parameters
has a marginal overlap with the much smaller 90\% C.L. of the $\nu_{\mu}$ oscillation parameters
(see the figure in page 11 of Ref.~\cite{Evans-HEP2009}),
the MINOS hint in favor of CPT violation is rather speculative.
Nevertheless,
it is interesting to notice that
a global separate analysis of neutrino and antineutrino data
in the framework of three-neutrino mixing with CPT violation
leads to different best-fit values of the oscillation parameters
of neutrinos and antineutrinos
with
$ \Delta\overline{m}^2_{\text{ATM}} \simeq \Delta\overline{m}^2_{\text{MINOS}} $
and
$ \sin^2\overline{\theta}_{\text{ATM}} \simeq \sin^2\overline{\theta}_{\text{MINOS}} $,
whereas
$ \Delta{m}^2_{\text{ATM}} \simeq \Delta{m}^2_{\text{MINOS}} $
and
$ \sin^2\theta_{\text{ATM}} \simeq \sin^2\theta_{\text{MINOS}} $
\cite{0908.2993}.
However,
in this paper we do not consider the MINOS hint in favor of CPT violation.
We concentrate our study on possible CPT violations in (V)SBL
$\nu_{e}$ and $\bar\nu_{e}$
disappearance due to squared-mass differences larger than about
$ 0.1 \, \text{eV}^2 $.

Besides those in Eqs.~(\ref{equ:pelel}) and (\ref{equ:paeae}),
it is possible to consider other, more complicated, expressions for
$P_{ee}$ and $P_{\bar e \bar e}$,
with additional energy-dependent terms in the oscillation phases which could be generated by
modified dispersion relations that are different for neutrinos and antineutrinos
(see, for example, Refs.~\cite{hep-ph/0309025,hep-ph/0506091,hep-ph/0606154,Hollenberg:2009tr,Esposito:2009ca,0907.1979}).
However, the introduction of more unknown parameters would make the analysis too cumbersome,
without much additional information on the potentiality of a neutrino factory experiment to test CPT invariance.
In fact, it is plausible that the additional energy-dependent terms in the oscillation phases
generate spectral distortions which would make the identification of new physics even easier
than in the simplest case that we consider.

In order to test CPT invariance (or {\em small} deviations from it) explicitly, it is convenient to define the averaged neutrino oscillation parameters
\begin{equation}
\theta \equiv \frac{1}{2} \left(  \theta_\nu + \theta_{\bar \nu} \right) \, , \quad \Delta m^2 \equiv \frac{1}{2} \left( \Delta m_\nu^2 + \Delta m_{\bar \nu}^2 \right)
\,,
\end{equation}
together with the CPT asymmetries
\begin{equation}
a_{\mathrm{CPT}} \equiv \frac{\theta_\nu - \theta_{\bar \nu}}{\theta_\nu + \theta_{\bar \nu}} \, , \quad
m_{\mathrm{CPT}} \equiv \frac{\Delta m_\nu^2 - \Delta m_{\bar\nu}^2}{\Delta m_\nu^2 + \Delta m_{\bar\nu}^2} \, ,
\label{asy}
\end{equation}
which are constrained in the range between $-1$ and $1$.
Then we have
\begin{eqnarray}
\theta_\nu  & = &  (1 + a_{\mathrm{CPT}}) \, \theta \, ,
\label{p1}
\\
\theta_{\bar \nu} & = & (1 - a_{\mathrm{CPT}}) \, \theta \, ,
\label{p2}
\\
\Delta m^2_\nu  & = &  (1 + m_{\mathrm{CPT}}) \, \Delta m^2 \,  ,
\label{p3}
\\
\Delta m^2_{\bar \nu} & = & (1 - m_{\mathrm{CPT}}) \,  \Delta m^2 \, .
\label{p4} 
\end{eqnarray}
The limit of CPT invariance (Eq.~(\ref{pee-cpt})) corresponds to $a_{\mathrm{CPT}} = m_{\mathrm{CPT}} = 0$.
In Section~\ref{cpt} we discuss the potentiality of neutrino factory experiments to discover
$ a_{\mathrm{CPT}} \neq 0 $
and/or
$ m_{\mathrm{CPT}} \neq 0 $.

The plan of the paper is:
in Section~\ref{ideal}
we define an ``ideal detector'' for the measurement of (V)SBL $\nu_{e}$ and $\bar\nu_{e}$ disappearance
at a neutrino factory, and we describe our treatment of geometric effects;
in Section~\ref{systematics}
we discuss the requirements for systematics;
in Section~\ref{disappearance} we discuss the sensitivity to (V)SBL $\nu_{e}$ and $\bar\nu_{e}$ disappearance
assuming CPT invariance,
with the survival probability in Eq.~(\ref{pee-cpt});
in Section~\ref{cpt} we discuss the sensitivity to CPT violation
considering the survival probabilities in Eqs.~(\ref{equ:pelel}) and (\ref{equ:paeae});
conclusions are presented in the final Section~\ref{conclusions}.

\section{Ideal detector and geometric effects}
\label{ideal}

\begin{figure}[t]
\begin{center}
\includegraphics[width=13cm]{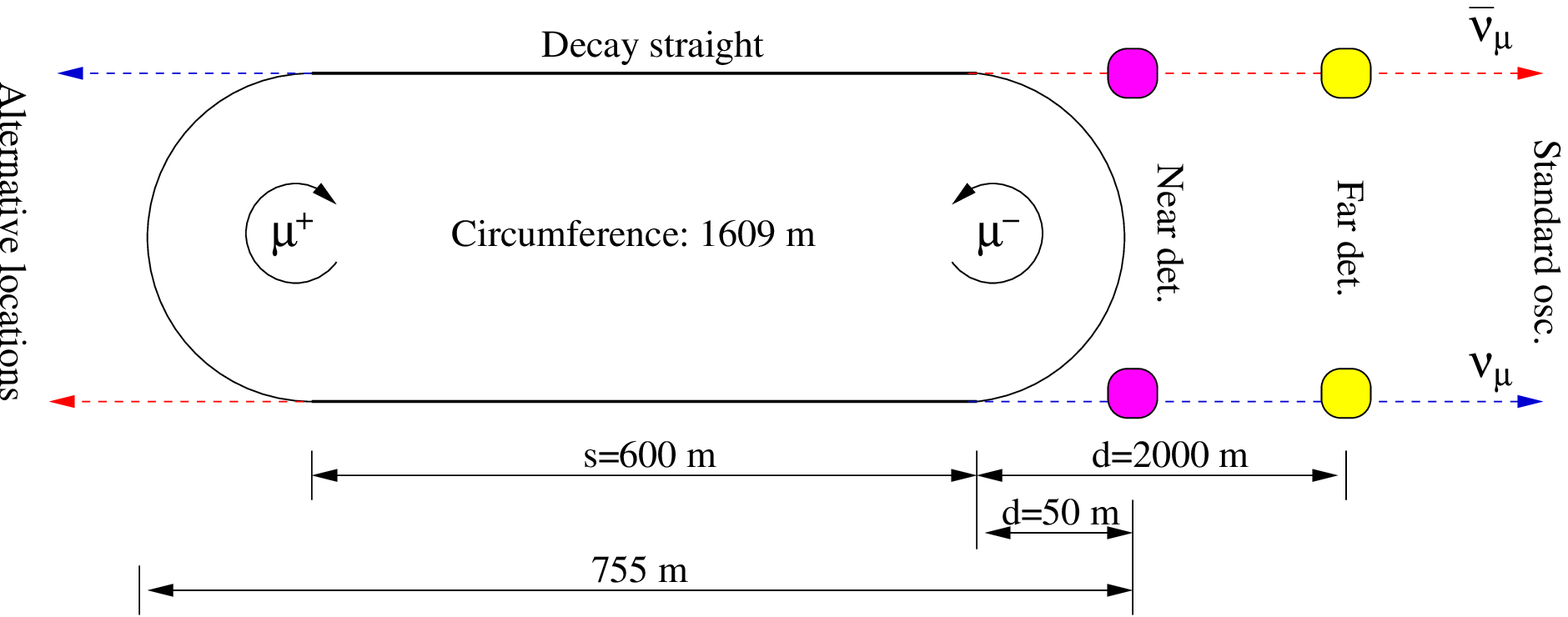}
\end{center}
\mycaption{\label{fig:ring} Geometry of the decay ring (not to scale). Two possible detector
locations are shown at $d=50 \, \mathrm{m}$ and $d=2 \, 000 \, \mathrm{m}$, where $d$ is the distance to the end of the decay straight. The baseline $L$ is the distance between production point and detector.}
\end{figure}

Our neutrino factory geometry is based on the International Design Study for the Neutrino Factory (IDS-NF) baseline setup~\cite{ids},
with the geometry illustrated in \figu{ring}.
We consider $2.5 \times 10^{20}$ useful muon decays per polarity and year, with muon energy $E_\mu=25 \, \mathrm{GeV}$.
For the total running time, we consider ten years.

In order to test SBL $\nu_e$ disappearance, we add detectors in front of the decay straights as illustrated in \figu{ring}.
Here ``near'' and ``far detectors'' refer to SBL $\nu_e$ disappearance only,
whereas the detectors for standard oscillations are much farther away and not relevant for our problem.
The straight sections are anticipated to be about
$s=600 \, \mathrm{m}$ long. The distance $d$ is the distance between the
end of the decay straight and the near detector.  The baseline $L$ is the
distance between production point and near detector, \ie, $d \le L \le d+s$. Since the $\mu^+$ and $\mu^-$ are assumed to circulate in different directions in the ring, we need pairs of detectors in front of the straights because we want to test CPT invariance.\footnote{Without CPT invariance test, detectors in front of one straight are sufficient. The detectors in front of the other straight only increase statistics then.} 

Since there are no specifications for near detectors at a neutrino factory yet (see \Ref~\cite{Abe:2007bi} for a generic discussion), we turn the argument around and formulate the requirements for the detectors for this measurement.
Our detectors are assumed to measure the total charged current rates with a 100\% detection efficiency; a lower efficiency will simply lead to a re-scaling of statistics and can be easily compensated by a larger detector mass.
The energy threshold is chosen to be $500 \, \mathrm{MeV}$,
similar to a Totally Active Scintillator Detector or an iron calorimeter,
and the energy resolution is taken as
\begin{equation}
\Delta E = \varepsilon \, \sqrt{ \frac{ E }{ E_{0} } }
\,,
\label{resolution}
\end{equation}
with $ \varepsilon = 0.55 \, \mathrm{GeV} $ and $ E_{0} = 1 \, \mathrm{GeV} $,
which is a conservative estimate for a magnetized iron calorimeter~\cite{ids}. Similarly, we assume that the neutral current level can be controlled at the level of $10^{-3}$ from all neutrinos in the beam (see, \eg, \Refs~\cite{Geer:2007kn,Bross:2007ts} in the context of a low energy neutrino factory).
However, we have tested that the results do not strongly depend on these three quantities.
We require an excellent flavor identification (at the level of $10^{-3}$ for the misidentification, as we will see later).
Charge identification is also desirable in order to reduce the contamination of the $\nu_{e}$ (or $\bar\nu_{e}$) signal by $\bar\nu_{e}$ (or $\nu_{e}$)
generated by possible (V)SBL $\bar\nu_\mu\to\bar\nu_e$ (or $\nu_\mu\to\nu_e$) oscillations.
However, we do not consider the backgrounds from charge misidentification explicitly.\footnote{The level of contamination depends on the oscillation model. Even for large mixing angles driving these oscillations
of the potential background, a charge misidentification level of about $10^{-3}$ would be sufficient.}
For the binning, we use 17~bins between $0.5$ and $25 \, \mathrm{GeV}$ with a bin size of $0.5 \, \mathrm{GeV}$ (1 bin) -- $1 \, \mathrm{GeV}$ (9 bins) -- $2 \, \mathrm{GeV}$ (5 bins) -- $2.5 \, \mathrm{GeV}$ (2 bins). 
As the main obstacles for the physics potential, we have identified the extension of the decay straights and the impact of systematics. We discuss the first issue below, and the second issue in the next section. Thereby, we define our ``ideal detectors'' as detectors with the above properties, but no backgrounds and systematics.

Our geometric treatment of the near detectors is based on \Ref~\cite{Tang:2009na},
which discusses the flux at near detectors in detail. Here we start from the differential event rate from a point source $dN_{\mathrm{PS}}/dE$ without oscillations.
Taking into account the extension of the straight and the geometry of the detector,
the averaged differential event rate is given by\footnote{Note that as a peculiarity compared to \Ref~\cite{Tang:2009na}, $dN_{\mathrm{PS}}/dE$ uses the unoscillated event rate, because the oscillation probability has to be integrated over.}
\begin{equation}
\frac{dN_{\mathrm{avg}}}{dE} = \frac{1}{s} \int\limits_{d}^{d+s}\frac{dN}{dE} dL = \frac{1}{s} \int\limits_{d}^{d+s}\frac{dN_{\mathrm{PS}}(L,E)}{dE} \, \varepsilon(L,E) \, P_{ee}(L,E) dL \, .
\end{equation}
Here $\varepsilon(L,E)=A_{\mathrm{eff}}/A_{\mathrm{Det}}$
parameterizes the integration over the detector geometry for a fixed baseline $L$ and given energy $E$
($A_{\mathrm{Det}}$ is the surface area of the detector
and $A_{\mathrm{eff}}$ is the effective surface area which takes into account the
angular dependence of the neutrino flux).
Since $dN_{\mathrm{PS}}/dE \propto 1/L^2$,
we can re-write this as
\begin{equation}
\frac{dN_{\mathrm{avg}}}{dE} = \frac{dN_{\mathrm{PS}}(L_{\mathrm{eff}},E)}{dE} \frac{L_{\mathrm{eff}}^2}{s} \int\limits_{d}^{d+s} \frac{ \varepsilon(L,E)}{L^2} \, P_{ee}(L,E) dL  =  \frac{dN_{\mathrm{PS}}(L_{\mathrm{eff}},E)}{dE}
\, \hat P(E)
\,, 
\label{equ:eavg}
\end{equation}
with the average efficiency ratio times probability\footnote{Note that \equ{eavg}
implies that in GLoBES a point source spectrum at the effective baseline $L_{\mathrm{eff}}$ can be used,
which has to be corrected by \equ{peff}. We perform \equ{peff} directly in the probability engine.}
\begin{equation}
\hat P(E)  \equiv \frac{L_{\mathrm{eff}}^2}{s} \int\limits_{d}^{d+s} \frac{ \varepsilon(L,E)}{L^2} \, P_{ee}(L,E) dL
\,,
\label{equ:peff}
\end{equation}
and the effective baseline
\begin{equation}
L_{\mathrm{eff}}=\sqrt{d (d+s) } \, ,
\label{effective-baseline}
\end{equation}
such that $\hat P(E) = 1$ for $\epsilon(L,E) \equiv P_{ee}(L,E)\equiv 1$.
We assume $\epsilon(L,E) \equiv 1$ (far distance approximation), which, to a good
approximation, is satisfied for ND4 of \Ref~\cite{Tang:2009na} (see Fig.~4 therein) for $d \gtrsim 50 \, \mathrm{m}$.
This detector is very small (200~kg), however, with a sufficient event rate. At a neutrino factory, the active volume of near detectors are probably going to be rather small, because high granularity and good track reconstruction will be more important than the active volume size~\cite{Abe:2007bi}. 
Our ``ideal'' test detectors therefore have 200~kg fiducial volume at very short distances. One can, for longer baselines, up-scale the detector mass as
\begin{equation}
m_{\mathrm{Det}} \simeq \frac{d \times (d+600 \, \mathrm{m})}{50 \, \mathrm{m} \times 650 \, \mathrm{m}} \, 0.2 \, \mathrm{t}
\label{equ:upscale}
\end{equation}
without strong geometric effects from the effective area of the detector (\ie, one still operates in the far distance limit).  However, one may choose a different technology for these larger detectors.

\begin{figure}[t]
\begin{center}
\includegraphics[width=10cm]{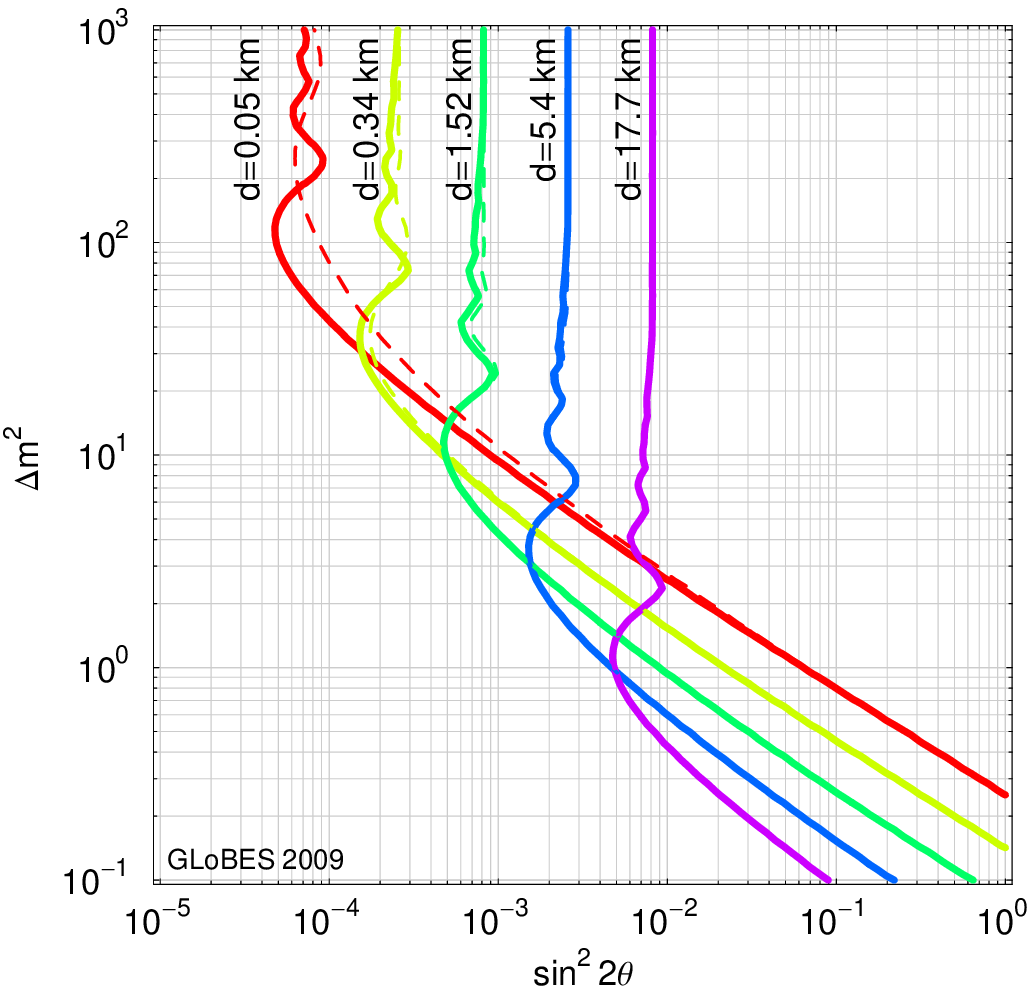}
\end{center}
\mycaption{\label{fig:es} Exclusion limit for several near detector distances $d$ and our ideal near detectors (CPT invariance assumed; 90\% CL, 2 d.o.f.; two near detectors in front of straights). The dashed curves illustrate the effect of including the averaging over the decay straight,
whereas the solid curves are without this averaging. The fiducial detector masses are fixed to 200~kg. Note that there is no systematics included in this figure. }
\end{figure}

For our simulation, we use the GLoBES software~\cite{Huber:2004ka,Huber:2007ji}. We define the exclusion limit as a function of $\sin^2 2 \theta$  and $\Delta m^2$ as the excluded region obtained in a $\chi^2$ analysis assuming a vanishing true value of $\theta$ (i.e. no oscillations).
In \figu{es}, we show this exclusion limit for several near detector distances including the effects of averaging over the decay straight (dashed curves) and without averaging (solid curves). This figure is based on our ``ideal'' detectors without taking into account systematics yet.
Obviously, the optimal detector locations depend on the region of sensitivity of $\Delta m^2$ which is of interest: the smaller $\Delta m^2$, the longer the baseline.
For instance, for $\Delta m^2 \simeq 1 \, \mathrm{eV}^2$, best sensitivity is obtained for $d \simeq 20 \, \mathrm{km}$, whereas for $\Delta m^2 \simeq 100 \, \mathrm{eV}^2$, a distance of the order $d=100 \, \mathrm{m}$ is optimal.  For short distances $d$ up to a few hundred meters, there is clearly an effect of the averaging over the decay straight. However, note that because of the $1/L^2$ weighting in \equ{eavg}, the effect becomes negligible for $d \gtrsim 1 \, \mathrm{km}$.
Compared to a classical beam dump experiment, one cannot get arbitrarily close to the source without loosing information.
In the next section, we will discuss the requirements for systematics.

We have also tested a low energy neutrino factory for this measurement, with similar success. However, in the absence of official numbers for the storage ring geometry and systematics, we will not discuss it in greater detail. In addition, note that the absolute performance is not {\em a priori} better than for a higher energy neutrino factory. For instance, assume that the distance $d$ is fixed for geometry reasons. Then the oscillation effect is, to a first approximation, proportional to $1/E^2$ (with $E$ the peak energy of the spectrum), but the statistics roughly increases as $E^3$ ($E^2$ from the beam collimation and $E$ from the cross sections), which means that the net effect is proportional to $E_\mu$. We observed this behavior in our simulation. 

\section{Requirements for systematics}
\label{systematics}

As far as systematics is concerned, it is well known from reactor experiments, such as
Double Chooz \cite{hep-ex/0606025} and Daya Bay \cite{hep-ex/0701029}, that electron neutrino disappearance is
most affected by the signal normalization uncertainty (see, \eg, \Refs~\cite{Huber:2003pm,Huber:2006vr}).
We expect the same for our measurement. However, compared to
reactor experiments, our signal normalization error does not mainly come from the knowledge on the flux, which
we may know to the level of 0.1\% using various mean monitoring devices~\cite{Abe:2007bi},
but from the knowledge of the cross sections. Because our neutrino energies span the
cross section regimes from quasi-elastic scattering, over resonant pion production, 
to deep inelastic scattering, it is not {\em a priori} simple to estimate the
accuracy of the cross sections knowledge at the time of the measurement. For reactor experiments,
on the other hand, the inverse beta decay cross sections are well known. Note that 
\Ref~\cite{0907.3145} also uses this well-understood detection reaction for a low-gamma beta beam,
whereas we will use a completely orthogonal strategy.

\begin{figure}[tp]
\begin{center}
\includegraphics[width=\textwidth]{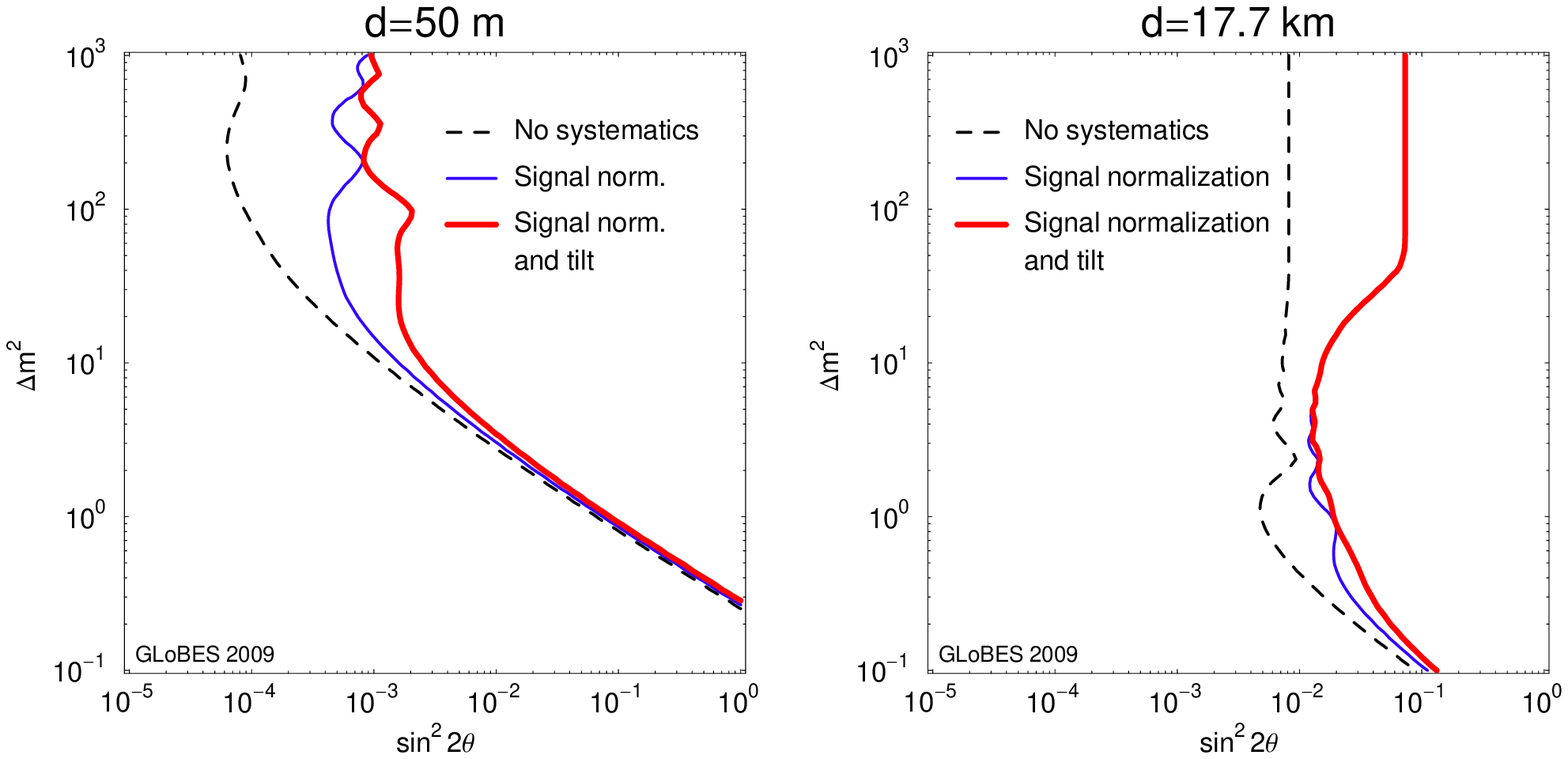}
\end{center}
\mycaption{\label{fig:sys} The effect of a different (hypothetical) systematical errors: A signal normalization error of 2.5\% and an additional spectral tilt error of 2.5\% has been applied to the exclusion limit for two different detector distances $d$ (90\% CL, 2 d.o.f.). The dashed curves refer to our ideal detectors, the solid curves include systematics. 
Here the fiducial mass is fixed to 200~kg, the effect of averaging over the decay straight is taken into account. Here CPT invariance is assumed.}
\end{figure}

Let us first of all illustrate what the main requirements for systematics are. As indicated above, we have tested in \figu{sys} the impact of a signal normalization error and an additional tilt error (tilting the shape of the spectrum). Although the errors are assumed to be rather optimistic (2.5\%), there is a significant impact on the sensitivities at all baselines, as we expected. 
Off the oscillation maxima, as visible in the right panel at large values of $\Delta m^2$ where $ P_{ee} \simeq 0.5 \, \sin^2 2 \theta $, the signal normalization error $\sigma_{\mathrm{Norm}}$ directly limits the sensitivity to $\sin^2 2 \theta \simeq 2 \, \sqrt{2.3} \, \sigma_{\mathrm{Norm}} \simeq 0.076 $ at $1 \sigma$ (2.3 is the $\Delta\chi^2$ corresponding to $1 \sigma$ for two degrees of freedom).
The tilt error tilts the spectrum linearly, and is a first order approximation for a shape error. It is especially important where the spectral information leads to a good sensitivity, in particular, for the shorter baselines (left panel). However, note that this (linear) tilt error cannot fully take into account the uncertainties in the cross sections, because the actual deviation may be non-linear. 
We have also tested the impact of backgrounds, energy resolution and energy threshold. The most important of these three systematics is the background, where the  sensitivity is basically limited by the product of background level and background uncertainty. Even for large uncertainties of the background, such as 20\%, this product limits the sensitivity to about $0.001 \times 20\% \simeq 10^{-4}$, which is beyond our expectations in the presence of a normalization uncertainty. 

In summary, the signal normalization and shape have to be either very well known, or very well measured.
The first requirement means that one needs very refined theoretical models for the cross sections, the second possibility means that one needs to measure the cross sections very well.
We follow the second approach by considering a setup with two sets of detectors
(\cf, \figu{ring}):
\begin{enumerate}
\item
 Near detectors at $d=50 \, \mathrm{m}$ with $m_{\mathrm{Det}}=200 \, \mathrm{kg}$.
\item
 Far detectors at $d=2 \, 000 \, \mathrm{m}$ with $m_{\mathrm{Det}}=32 \, \mathrm{t}$.
\end{enumerate}
The signal measured with the near detectors fixes the normalization and shape of the unoscillated signal
(for small enough $\Delta m^2$).
The far detectors are up-scaled versions of the near detectors following \equ{upscale}, which means that
geometric effects are almost negligible.
The near detectors have optimal sensitivity at a few hundred $\mathrm{eV}^2$ (VSBL),
whereas the far detectors have optimal sensitivity at a few $\mathrm{eV}^2$ (SBL).
Note that longer baselines may be even better for the far detectors, but then the depth difference between storage ring and detectors may become unrealistically large. On the other hand, for distances much shorter than 2~km, one significantly looses sensitivity for small $\Delta m^2$.

For systematics, we adopt the most conservative point of view, \ie, we assume that we hardly know anything about the cross sections, neither the normalization nor the shape, but that the cross sections are fully correlated among all detectors measuring the cross sections. Such an error is often called ``shape error'' and is uncorrelated among the bins. 
In summary, we include the following systematical errors similar to the reactor experiments in \Ref~\cite{Huber:2006vr}, and we have tested their impact (we have switched off systematical errors to test their impact):
\begin{description}
 \item[Shape errors] uncorrelated among bins and $\nu$-$\bar \nu$, but fully correlated among the detectors. These errors include cross section errors, scintillator or detector material properties, \etc. In addition, flux errors can be included here (the detectors only measure the product of flux and cross section for the disappearance channel). We estimate this error to be 10\%. However, even a larger error does not matter if both near and far detectors are present, but only errors considerably smaller than $10^{-3}$ improve the result significantly (which is absolutely unrealistic for this type of systematics).
\item[Normalization errors] uncorrelated between the near and far detectors. These relative normalization errors
come from the knowledge on fiducial mass, detector normalization, and analysis cuts (uncorrelated between the detectors). They are typically small if similar detectors are used. For reactor experiments (Double Chooz \cite{hep-ex/0606025}), this error is about 0.6\%, which we use as an estimate. We have tested that there is little dependence on this error unless it can be reduced to the level of $10^{-4}$ (then there is a small improvement), if the other systematics is present.
\item[Energy calibration errors] uncorrelated between the near and far detectors of the order 0.5\% are used (similar to the reactor experiments). As we have tested, they are of secondary importance if all the other systematics is present.
\item[Backgrounds] at the level of $10^{-3}$ from neutral current events \etc\ are assumed, known to the level of 20\%
(somewhat conservative estimate from a neutrino factory). If all the other errors are present, backgrounds hardly matter.
\end{description}

\begin{figure}[t!]
\begin{center}
\includegraphics[width=\textwidth]{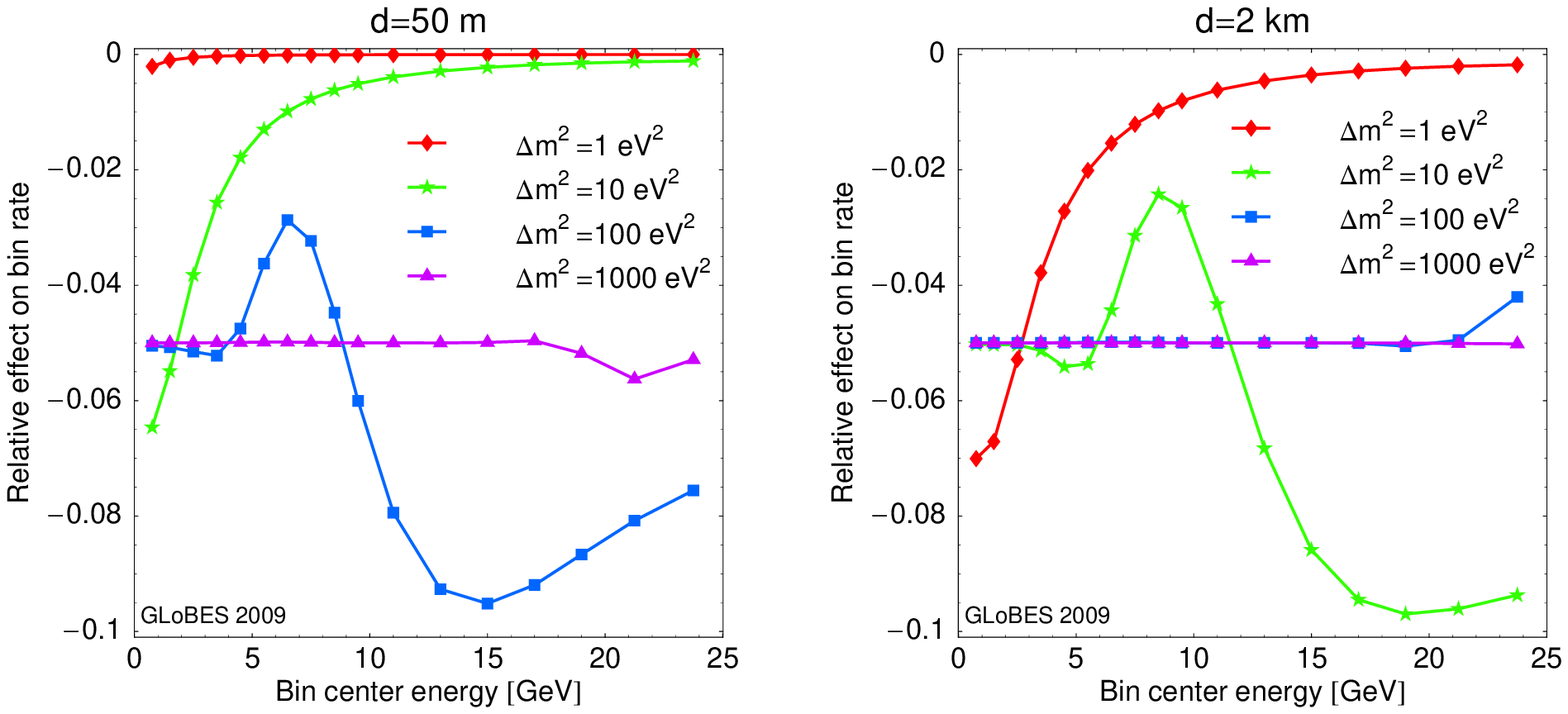}
\end{center}
\mycaption{\label{fig:events}
Relative effect on the binned (neutrino) event rates for
several values of $ \Delta m^2 $, and $ \sin^2 2 \theta = 0.1 $,
in the near (left) and far (right) detectors.
For each energy bin we plotted
$ (R-R_{0})/R_{0} $
where $R$ and $R_{0}$ are the expected rates with and without oscillations.
}
\end{figure}

The effect of electron neutrino disappearance on the event rates of the individual bins 
is illustrated in \figu{events} for the near (left) and far (right) detector for several values of $\Delta m^2$. For relatively small $\Delta m^2 \sim 1 \, \mathrm{eV}^2$ (diamond curves), the near-far combination will perform similar to the reactor experiments with two detectors, where the near detector measures the shape and the far detector the oscillation effect. For $\Delta m^2 \gg 1000 \, \mathrm{eV}^2$ (\cf, triangle curves for comparison), the oscillations average out in both detectors, and $\sin^2 2 \theta$ can only be constrained to the level of the shape errors (whereas $\Delta m^2$ cannot be measured). For $\Delta m^2 \sim 100 \, \mathrm{eV}^2$ (box curves), the oscillation effect will mainly take place in the near detector, whereas the far detector measures the shape (after averaging). For $\Delta m^2 \sim 10 \, \mathrm{eV}^2$ (star curves), the situation is most complicated: there are oscillation effects in both detectors, which can lead to intricate parameter correlations.

\section{Results for CPT invariance}
\label{disappearance}

All results presented in this section are based on our two-baseline setup without refined systematics treatment,
assuming CPT invariance, \ie, the equal electron neutrino and antineutrino survival probabilities in Eq.~(\ref{pee-cpt}).

\begin{figure}[t!]
\begin{center}
\includegraphics[width=11cm]{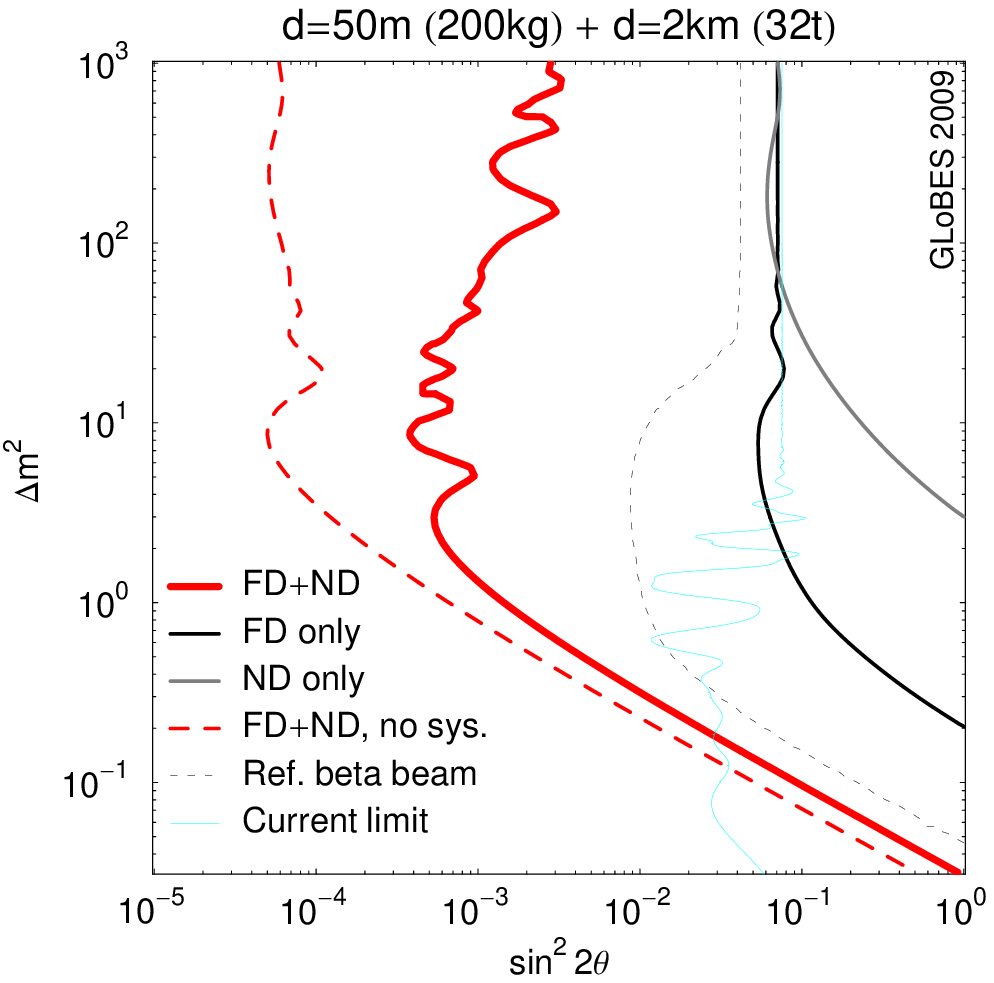}
\end{center}
\mycaption{\label{fig:fnsys} Exclusion limit in the $\sin^2 2 \theta$-$\Delta m^2$ plane for our default configuration including systematics (thick solid curve, 90\% CL, 2 d.o.f.).
The thick dashed curve refers to our ideal detectors (no systematics), with near (ND) and far (FD) detectors combined.  The thin solid curves illustrate the results for the near (50~m) and far (2~km) detectors if operated separately, but with full systematics. The effects of averaging over the decay straights are taken into account. The thin dashed curve corresponds to the default beta beam setup from \Ref~\cite{0907.3145} for comparison.
The thin gray/cyan curve is the current limit from Bugey\protect\cite{Declais:1995su} + Chooz\protect\cite{hep-ex/0301017} (taken from \Ref~\cite{Acero:2007su}). }
\end{figure}

\figu{fnsys} shows the performance of our near-far model (thick curve), where the effect of using only one set of detectors (near or far) is shown separately as thin curves. If only one set of detectors is used, the result will be limited by the 10\% shape errors, \ie, it depends on the assumptions used. However, if the two sets of detectors are used, the impact of systematics cancels and the result is very robust with respect to the assumptions.  From the above discussion, it should be clear that the results in this case do not depend very much on the actual numbers for the systematical errors.
Nevertheless there is a considerable deviation from the no-systematics case (dashed curve). The improvement towards this hypothetical sensitivity requires a very good understanding of the cross sections at the level of the $\sin^2 2 \theta$ sensitivity. We have also checked that the performance cannot even be significantly improved with considerably larger detectors, because of the systematics limitation (even without the geometric effect of the beam included). 

Figure~\ref{fig:fnsys} shows that the sensitivity of a neutrino factory experiment to (V)SBL $\nu_{e}$ disappearance
represents a dramatic improvement with respect to the sensitivity of reactor experiments,
which is at the level of $ \sin^2 2 \theta \sim 10^{-1} $ at large values of $ \Delta m^2 $ (\cf, thin gray/cyan curve).
Moreover, the neutrino factory measurement with the near-far detector setup discussed in Section~\ref{systematics}
is model-independent,
whereas reactor measurements of $P_{\bar e \bar e}$ depend on the calculated flux of $\bar\nu_{e}$'s produced in a reactor.
Reactor neutrino experiments cannot take advantage of the near-far detector approach to get a model-independent result
for (V)SBL $\nu_{e}$ disappearance,
because for a typical reactor neutrino energy of 1~MeV
the oscillation length corresponding to $ \Delta m^2 \approx 10^2 \, \mathrm{eV}^2 $ is of the order of 1~cm.

It is interesting to note that the near-far detector setup that we have chosen
is sensitive to $\nu_{e}$ disappearance with small mixing
($ \sin^2 2 \theta \gtrsim 2 \times 10^{-3} $)
for values of $ \Delta m^2 $ as large as $ 10^{3} \, \mathrm{eV} $.
The condition for the observation of a spectral distortion
caused by neutrino oscillations is that the uncertainty of the phase of the oscillations
due to the energy resolution in Eq.~(\ref{resolution}) is smaller than about $\pi/2$.
One can easily find that this happens for neutrino energies
\begin{equation}
E \gtrsim
\left[
\frac{ \varepsilon \, \Delta m^2 \, L_{\mathrm{eff}} }{ 2 \pi \, E_{0}^{1/2} }
\right]^{2/3}
\,,
\label{emin}
\end{equation}
where we have considered the effective baseline in Eq.~(\ref{effective-baseline}).
Since for the near detector $ L_{\mathrm{eff}} \simeq 180 \, \mathrm{m} $,
if $ \Delta m^2 = 10^{3} \, \mathrm{eV} $ the condition (\ref{emin})
is satisfied for
$ E \gtrsim 18 \, \mathrm{GeV} $.
Since for the assumed $E_{\mu}=25\,\mathrm{GeV}$ the neutrino energy spectrum extends 
up to $25\,\mathrm{GeV}$,
as shown by the curve in Fig.1 of Ref.~\cite{Tang:2009na}
with off-axis angle $\theta=0^{\circ}$,
the oscillations are not completely averaged out in the highest-energy bins.
This is illustrated in the left panel of Fig.~\ref{fig:events},
in which the line corresponding to $ \Delta m^2 = 10^{3} \, \mathrm{eV} $
has the constant averaged value $ 0.5 \, \sin^2 2 \theta - 1 = - 0.05 $
(for the assumed $ \sin^2 2 \theta = 0.1 $)
only for $ E \lesssim 10 \, \text{GeV} $.
Other curves illustrate the distortion of the event rate spectrum for
smaller values of $ \Delta m^2 $.
One can see that the lower limit of the sensitivity to $ \Delta m^2 $ of the near detector
is about $ 1 \, \mathrm{eV}^2 $,
which instead produces a strong spectral distortion in the far detector
(right panel of Fig.~\ref{fig:events}).

We also show in \figu{fnsys} a comparison with the default setup in \Ref~\cite{0907.3145} (thin dashed curve).
This setup uses a low-gamma ($\gamma \simeq 30$) beta beam using inverse beta decay as detection interaction,
which means that it is not surprising that our result is about an order of magnitude better. Compared to \Ref~\cite{0907.3145}, which uses only one detector and therefore runs in the systematics limitation in the larger $\Delta m^2$ range, we also have very good sensitivity for large $\Delta m^2$. While both approaches rely on near detectors receiving neutrinos from a storage ring, they are conceptually very different: \Ref~\cite{0907.3145} uses the fact that the inverse beta decay reaction is well known to control systematics, whereas we control the shape error with two sets of detectors in the fashion of the new generation of reactor experiments.

\begin{figure}[t!]
\begin{center}
\includegraphics[width=\textwidth]{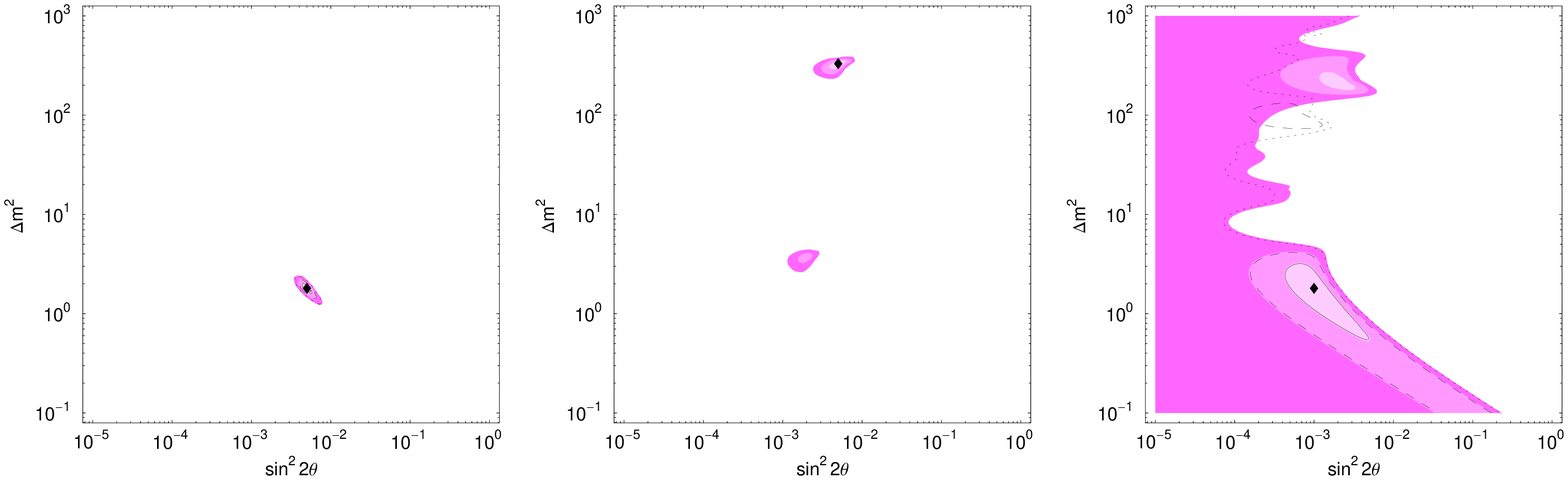}
\end{center}
\mycaption{\label{fig:fitscpt} Fits in the $\sin^2 2 \theta$-$\Delta m^2$ plane for three chosen test-points marked by the diamonds ($1\sigma$, $2\sigma$, $3 \sigma$, 2 d.o.f.). 
Here CPT invariance is assumed. Near (50~m) and far (2~km) detectors are used with our systematics model, the effects of averaging over the decay straights are taken into account. The unshaded contours show the result without averaging effects over the straights. They are too small to be visible in the middle panel.}
\end{figure}

It is interesting to examine not only the sensitivity
of our experimental setup to (V)SBL $\nu_{e}$ disappearance,
which corresponds to a negative result producing an exclusion curve as that in \figu{fnsys},
but also what could be the results if a signal is observed, \ie, $\nu_e$ and $\bar\nu_e$ disappear.

In \figu{fitscpt}, we show three qualitatively different possible results for the test values of the neutrino oscillation parameters marked by the diamonds.
In the left panel, no degenerate solutions are present, and the parameters can be very well measured. There is hardly an effect of the averaging over the decay straights, as one can read off from the differences between the shaded and unshaded contours, because the far detectors dominate the sensitivity and oscillations have not yet developed at the near detectors. In the middle panel, we still have an excellent measurement dominated by the near detectors. In this case, however, the averaging effects over the straights are very important, and the contours without averaging are hardly visible. In particular, a degenerate solution appears at a smaller $\Delta m^2$. In the right panel, we show an even more extreme case, where only at the $2 \sigma$ confidence level $\sin^2 2 \theta=0$ can be excluded.

\section{CPT violation}
\label{cpt}

\begin{figure}[tp]
\begin{center}
\includegraphics[height=0.8\textheight]{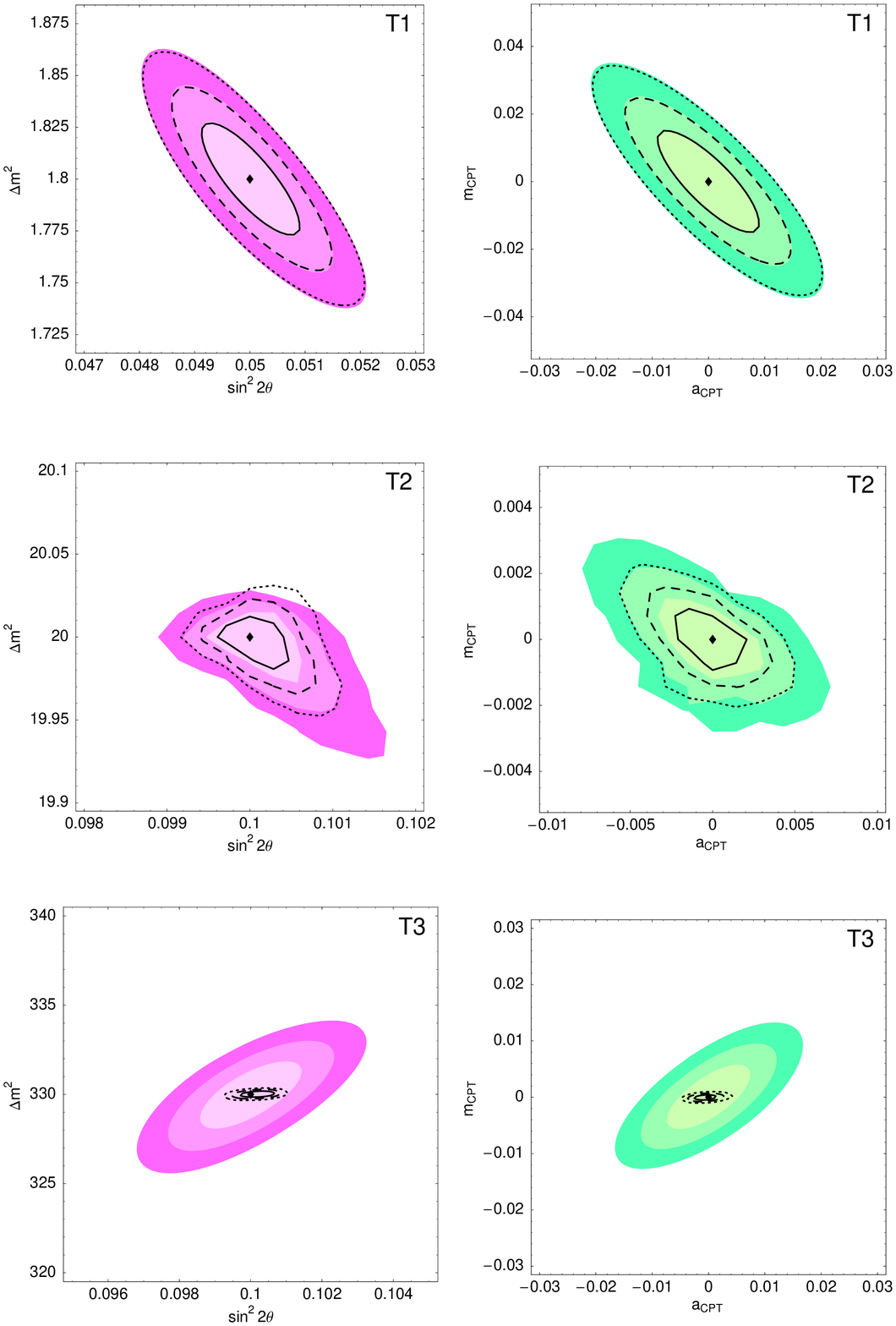}
\end{center}
\mycaption{\label{fig:fits} Best-fit regions in the $\Delta m^2$-$\sin^2 2 \theta$ and $a_{\mathrm{CPT}}$-$m_{\mathrm{CPT}}$ planes for the test points defined in the main text ($1\sigma$, $2 \sigma$, $3 \sigma$, 2 d.o.f.). 
Note that different baselines where chosen for the test points. The dashed curves represent the results without taking into account the averaging over the decay straights. }
\end{figure}

In this section we discuss the potentiality of the experimental setup described in \figu{ring},
with two pairs of near-far detectors,
to reveal a violation of CPT symmetry,
considering the different electron neutrino and antineutrino survival probabilities in Eqs.~(\ref{equ:pelel}) and (\ref{equ:paeae})
as functions of the CPT asymmetries in Eq.~(\ref{asy}).

Since there are four independent parameters,
given by Eqs.~(\ref{p1})--(\ref{p4}),
for simplicity we consider three test points inspired by \Refs~\cite{Acero:2007su,Giunti:2007xv,Giunti:2009zz}:
\begin{eqnarray}
\mathrm{T1:} & \quad & \sin^2 2 \theta= 0.05 \, , \quad \Delta m^2 = 1.8 \, \mathrm{eV}^2  \, ,  \\
\mathrm{T2:} & \quad & \sin^2 2 \theta= 0.1 \, , \quad \Delta m^2 = 20 \, \mathrm{eV}^2 \, ,  \\
\mathrm{T3:} & \quad & \sin^2 2 \theta= 0.1 \, , \quad \Delta m^2 = 330 \, \mathrm{eV}^2 \, ,
\end{eqnarray}
and $a_{\mathrm{CPT}}=m_{\mathrm{CPT}}=0$.
We fit the corresponding simulated data allowing for non-zero values of $a_{\mathrm{CPT}}$ and $m_{\mathrm{CPT}}$
in order to explore the sensitivity to the measurement of these parameters.

The test point T1 is motivated by the best-fit of the data of the Bugey SBL reactor experiment \cite{Declais:1995su},
which is compatible with the data of the Chooz reactor experiment \cite{hep-ex/0301017}
and the neutrino oscillation explanation of the Gallium anomaly
\cite{Acero:2007su}.
The test points T2 and T3 are motivated by a possible explanation
the MiniBooNE low-energy anomaly through VSBL $\nu_e$ disappearance
which is compatible with the neutrino oscillation explanation of the Gallium anomaly \cite{Giunti:2007xv,Giunti:2009zz}.
Even if values of $ \Delta m^2 $ larger than about $ 1 \, \text{eV}^2 $
are incompatible with the existing standard cosmological bound on the sum of neutrino masses
\cite{0805.2517,0809.1095},
we think that it is wise to test such bound in laboratory experiments.
A violation of the bound may lead to a discovery of fundamental new physics related to
non-standard cosmological effects.

The best-fit regions for the three test points are shown in \figu{fits}. The dashed curves represent the results without taking into account the averaging over the decay straights. The test-point T1 (upper row), with a relatively small $\Delta m^2$, is dominated by the far detectors, whereas in the near detectors (almost) no oscillations are present. Therefore, the cross sections can be directly reconstructed from the near detectors, and the fits are very clean. The effects of averaging over the straights are small because the signal sits in the far detectors, which sees a point source. The oscillation parameters can be measured at the level of 2\% ($1\sigma$), and the CPT invariance can be constrained at the same level.

The test-point T3 (lower row of \figu{fits}), is dominated by the short baseline, which means that the averaging effects over the straights are very important. The longer baseline measures the product of cross sections and $1- 0.5 \, \sin^2 2 \theta$, which means that $\Delta m^2$ can, before the averaging over the straights (dashed curves), be very well measured compared to the mixing angle since it remains as a net effect between the two detectors. Only after the averaging effects, both oscillation parameters can be measured at the level of 1\% ($1\sigma$), and the CPT invariance can be constrained at a similar level.

The test-point T2 (middle row of \figu{fits}) shows a complicated case with an intricate interplay between systematics and oscillation parameter correlations. Since there is an oscillation effect in both baselines, this case does not correspond to a classical near-far detector combination. The a priori excellent precisions for the oscillation parameters are spoilt by some complicated correlations. Nevertheless, percent level precisions are possible.

\begin{figure}[tp]
\begin{center}
\includegraphics[width=\textwidth]{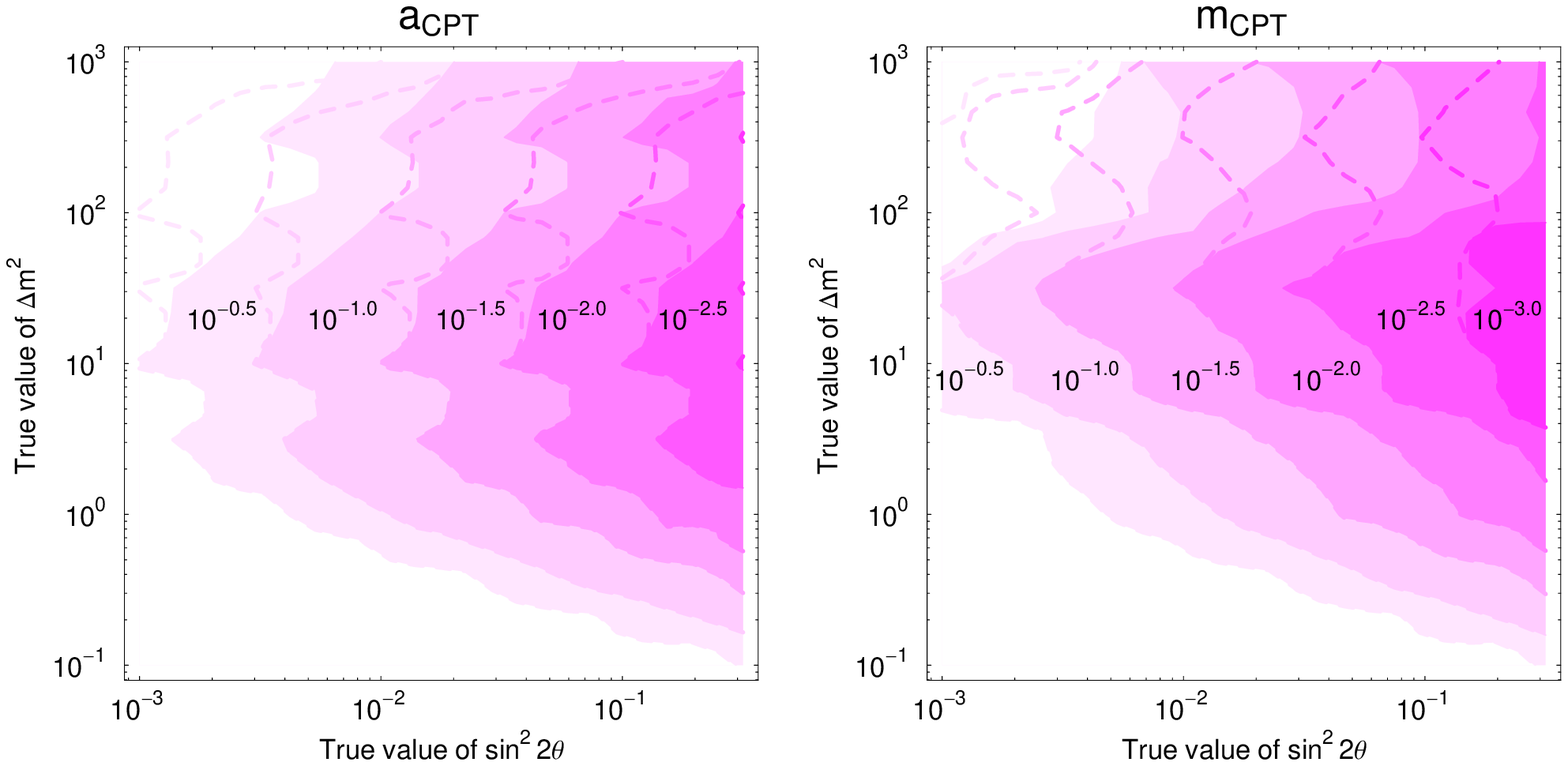}
\end{center}
\mycaption{\label{fig:cpt} Discovery reach for CPT violation from $a_{\mathrm{CPT}}$ (left panel) or $m_{\mathrm{CPT}}$ (right panel) as a function of the true $\sin^2 2 \theta$ and true $\Delta m^2$. The different contours indicate for how small (true) values of $a_{\mathrm{CPT}}>0$ (left) or $m_{\mathrm{CPT}}>0$ (right) CPT violation will discovered at the $3 \sigma$ confidence level, as labeled at the contours. The dashed curves show the result if the averaging over the decay straights is not taken into account.}
\end{figure}

Instead of constraining CPT invariance, we can also discuss the discovery reach for CPT violation. In this case, we assume that nature has implemented a small (positive) $a_{\mathrm{CPT}}$  or $m_{\mathrm{CPT}}$, and we fit
the simulated data with the fixed parameters $a_{\mathrm{CPT}}=m_{\mathrm{CPT}}=0$ (corresponding to CPT invariance),
while we  marginalize over the oscillation parameters $\sin^2 2 \theta$ and $\Delta m^2$. We show in \figu{cpt} the discovery reach for CPT violation from $a_{\mathrm{CPT}}$ (left panel) or $m_{\mathrm{CPT}}$ (right panel) as a function of the true $\sin^2 2 \theta$ and true $\Delta m^2$. The different contours indicate for how small (true) values of $a_{\mathrm{CPT}}>0$ (left) or $m_{\mathrm{CPT}}>0$ (right) CPT violation will discovered at the $3 \sigma$ confidence level, as labeled at the contours. The dashed curves show the result if the averaging over the decay straights is not taken into account.

From \figu{cpt}, CPT violation may be discovered even if it is as small as $10^{-3}$, provided that $\sin^2 2 \theta$ is large enough. However, even for very small $\sin^2 2 \theta$, a CPT violation of order unity is testable with our setup.  Note that for larger $\Delta m^2$ and especially for $m_{\mathrm{CPT}}$, the averaging over the decay straights strongly reduces the performance (by about one order of magnitude).

\begin{figure}[tp]
\begin{center}
\includegraphics[width=0.5\textwidth]{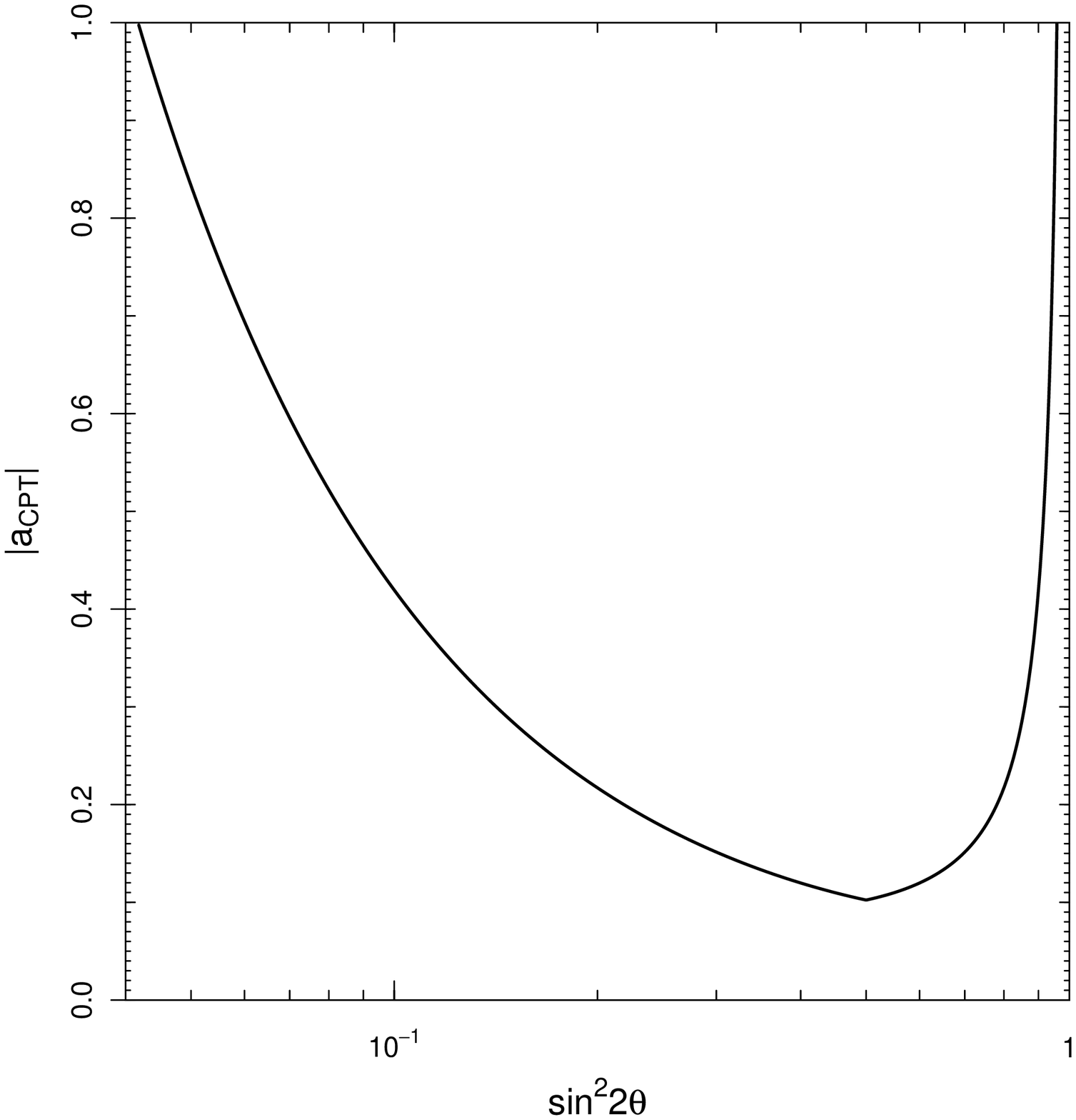}
\end{center}
\mycaption{\label{fig:acpt}
Lower limit for $|a_{\mathrm{CPT}}|$
obtained from Eq.~(\ref{acpt1}) with
$ A_{ee}^{\mathrm{CPT}} < - 0.08 $,
which is the 95\% C.L. limit found in \Ref~\cite{Giunti:2009zz}.
}
\end{figure}
In \Ref~\cite{Giunti:2009zz}, a difference of
$ A_{ee}^{\mathrm{CPT}} \equiv P_{ee} - P_{\bar e \bar e}= - 0.17 {}^{+0.09}_{-0.07} $
at 90\% C.L. was identified as the
asymmetry between the electron neutrino and anti-neutrino
VSBL disappearance probabilities which can explain
the Gallium radioactive source experiments anomaly \cite{Abdurashitov:2009tn}
and
the MiniBooNE low-energy anomaly \cite{AguilarArevalo:2008rc}
without conflicting with the absence of $\bar\nu_{e}$ disappearance in
reactor neutrino experiments
(see Ref.~\cite{hep-ph/0107277}).
It is interesting to investigate if such CPT violation can be measured
in a neutrino factory experiment with the near-far pairs of detectors that we have considered so far.

Since in \Ref~\cite{Giunti:2009zz} $\Delta m^2$ was considered to be large,
in the range
\begin{equation}
20 \, \text{eV}^{2}
\lesssim
\Delta{m}^{2}
\lesssim
330 \, \text{eV}^{2}
\,,
\label{dm2}
\end{equation}
the neutrino and antineutrino survival probabilities were assumed to be averaged,
leading to
$ A_{ee}^{\mathrm{CPT}} = 0.5 \left( \sin^2 2\theta_{\bar\nu} - \sin^2 2\theta_{\nu} \right) $.
In this case,
the asymmetry $a_{\mathrm{CPT}}$ is given by
\begin{equation}
a_{\mathrm{CPT}}
=
\frac{ 1 }{ 4 \theta }
\,
\arcsin\left( \frac{ - 2 A_{ee}^{\mathrm{CPT}} }{ \sin 4 \theta } \right)
\,.
\label{acpt1}
\end{equation}
Since
$ |a_{\mathrm{CPT}}| \leq 1 $,
the mixing angle has a lower limit which depends on the value of $A_{ee}^{\mathrm{CPT}}$.
Moreover,
since $ | \sin 4 \theta | \leq 1 $
and $ \theta \leq \pi/2 $,
also
$ |a_{\mathrm{CPT}}| $ has a lower limit,
which is plotted in Fig.~\ref{fig:acpt} for
$ A_{ee}^{\mathrm{CPT}} < - 0.08 $,
which is the 95\% C.L. limit found in \Ref~\cite{Giunti:2009zz}.
One can see that the bound on $A_{ee}^{\mathrm{CPT}}$
implies that
$ \sin^22\theta \gtrsim 4 \times 10^{-2} $
and
$ |a_{\mathrm{CPT}}| \gtrsim 0.10 $.
Confronting these values with the left panel in Fig.~\ref{fig:cpt},
and taking into account that we consider the large values of
$\Delta{m}^{2}$
in the range (\ref{dm2}),
it is clear that the CPT violation required by $ A_{ee}^{\mathrm{CPT}} \lesssim - 0.08 $
will be easily discovered
in a neutrino factory experiment with the near-far pairs of detectors that we have considered.

\section{Summary and conclusions}
\label{conclusions}

In this work we have discussed the potentiality of testing
Short-BaseLine (SBL, with $ 10^{-1} \lesssim \Delta{m}^{2}_{\text{SBL}} \lesssim 10 \, \text{eV}^2 $) and
Very-Short-BaseLine (VSBL, with $ 10 \lesssim \Delta{m}^{2}_{\text{VSBL}} \lesssim 10^3 \, \text{eV}^2 $)
electron neutrino disappearance in a neutrino factory experiment,
based on the current setup of the International Design Study for the Neutrino Factory (IDS-NF)~\cite{ids}.
Since this setup uses both muon and anti-muon decays, a possible difference between the neutrino and antineutrino disappearance  can be studied, which could constitute a revolutionary discovery of CPT violation.

We showed that for these purposes the ideal configuration would be
the two pairs of near-far detectors (shown in Fig.~\ref{fig:ring})
in a similar fashion to reactor experiments with near and far
detectors (Double Chooz \cite{hep-ex/0606025}, Daya Bay \cite{hep-ex/0701029}, \etc) to cancel systematics.
The near detectors are chosen to be at a distance of about 50~m
from the muon storage ring, in order to be sensitive to oscillations
due to a $\Delta m^2$ as large as about $10^3 \, \mathrm{eV}^2$.
For the far detectors an appropriate distance
from the muon storage ring is about 2~km, which gives a good sensitivity
to oscillations
generated by a $\Delta m^2$ as small as about $10^{-1} \, \mathrm{eV}^2$.
In this way,
it is possible to explore (V)SBL
$\nu_{e}$ and $\bar\nu_{e}$ disappearance
with effective oscillation amplitude
$\sin^22\theta$ as small as about $10^{-3}$
for
$\Delta m^2 \gtrsim 1 \, \mathrm{eV}^2$
(see Fig.~\ref{fig:fnsys})
taking advantage of the comparison of the event rates
measured in the near and far detectors,
which reduces dramatically the systematic uncertainties due to
insufficient knowledge of the cross sections
(see the discussion in Section~\ref{systematics}).

We have also shown, in Section~\ref{cpt}, that the chosen detector setup
provides a good sensitivity to the measurement of
a difference of the rates of $\nu_{e}$ and $\bar\nu_{e}$ disappearance
which would be a signal of CPT violation. For instance, our setup is
sensitive to an asymmetry between the neutrino and antineutrino
mass squared differences at the level of up to $10^{-3}$, depending
on the value of the mixing angle.
Let us emphasize that a discovery of CPT violation would represent a revolution in our
knowledge of fundamental physics,
because the CPT symmetry
is a fundamental symmetry of local relativistic Quantum Field Theory.
Therefore,
pursuing this line of investigation is of fundamental importance.

\subsubsection*{Acknowledgments}

We would like to thank Sanjib Agarwalla for providing the beta beam reference
curve in \figu{fnsys}, and Patrick Huber for useful discussions.

This work was supported by the European Union under the European Commission
Framework Programme~07 Design Study EUROnu, Project 212372. W. Winter also would like to acknowledge support
from the Emmy Noether program of Deutsche Forschungsgemeinschaft.

C. Giunti would like to thank the Department of Theoretical Physics of the University of Torino
for hospitality and support.


\end{document}